# Mechanisms of soot-aggregate restructuring and compaction

**Joel C. Corbin**[1,2]**, Robin L. Modini**[1]**, and Martin Gysel-Beer**[1]

[1]Laboratory of Atmospheric Chemistry, Paul Scherrer Institute, 5232 Villigen, Switzerland

[2]Now at: Metrology Research Centre, National Research Council Canada, 1200 Montreal Rd, Ottawa, Canada

*Correspondence to:* Joel.Corbin@nrc-cnrc.gc.ca

**Abstract.**

Soot aggregates form as open, fractal-like structures, but aged atmospheric particles are often observed to be restructured into more compact shapes. This compaction has a major effect on the radiative properties of the aggregates, and may also influence their aerosol-cloud interactions, toxicity, and deposition in human lungs. Recent laboratory studies have presented conflicting arguments on whether this compaction occurs during condensation or during evaporation. In this three-part study, we combine theory and experiments to explain these conflicting results. First, we review the surface-science literature and identify explicit mechanisms for condensation-compaction as well as evaporation-compaction. We also identify a mechanism for *avoiding* compacting during condensation, based on heterogeneous nucleation theory and the kinetic barriers to capillary formation. Second, we review the soot-restructuring literature and find evidence for all of the identified compaction mechanisms, the most common being condensation-compaction. Some atmospheric studies have reported non-compacted soot internally mixed with other material. We attribute these statistically rare observations to coagulation, which is a less-common mode of soot processing than condensation. Third, we present new experimental results from a study in our laboratory where the surface tension of anthracene coatings was "switched on" or "switched off" by using solid or liquid phases during addition and removal. Consequently, we demonstrate condensation compaction, evaporation compaction, and no compaction, for the same soot source. Overall, our study indicates that it is most reasonable to assume that the majority of soot particles will undergo compaction when coated by condensing vapours in combustion systems and in the atmosphere. Compaction might be avoided by those soot particles which become coated by coagulation rather than condensation.

## 1 Introduction

Globally, soot particles are predicted to absorb comparable amounts of solar radiation as carbon dioxide, with the exact absorption amount depending on the concentration of soot particles and their mixing with other particulate material (PM) (Ramanathan and Carmichael, 2008; Bond *et al.*, 2013). However, describing the radiative properties of soot BC as a function of atmospheric lifetime remains a major challenge (Cappa *et al.*, 2012; Jacobson, 2013; Cappa *et al.*, 2013) due to variability in the detailed morphologies of the BC spherules which comprise soot (Yon *et al.*, 2014; Radney *et al.*, 2014) and the various possible configurations of soot with other internally-mixed material. A key issue is whether the spherules comprising soot aggregates are arranged in extremely open, fractal-like structures (Eggersdorfer and Pratsinis, 2013) or more compact, spheroidal shapes (Zangmeister *et al.*, 2014). This change of structure strongly influences their optical properties (Qiu *et al.*, 2012; Scarnato *et al.*, 2013; Radney *et al.*, 2014).

The open and compact limits of soot spherule morphology are defined by the processes of aggregate formation and of atmospheric aging. Soot aggregates, like other flame-synthesized aggregates, are formed by the diffusion-limited clustering of diameter roughly $10 - 80$ nm spherules (Eggersdorfer and Pratsinis, 2013; **?**), as observed at or near combustion source. However, electron microscopy of atmospherically aged soot often shows these spherules to be highly compacted (Xiong and Friedlander, 2001; Abel *et al.*, 2003; Shi *et al.*, 2003; Johnson *et al.*, 2005; Niemi *et al.*, 2006; Adachi and Buseck, 2008; Worringen *et al.*, 2008; Adachi *et al.*, 2010; Coz and Leck, 2011; Adachi and Buseck, 2013; China *et al.*, 2014, 2015; Bhandari *et al.*, 2019), apparently due to internal mixing with ("coating by") non-refractory material.

Historically, it has been widely believed that soot particles were compacted by surface tension forces upon the condensation of coatings (Kütz and Schmidt-Ott, 1992; Zhang *et al.*, 2008; Qiu *et al.*, 2012; Khalizov *et al.*, 2013) (see also Section 3). This belief generally relied on indirect evidence (Section 3), due to the difficulty of measuring the structure of a soot particle while it remains coated. In contrast to this common belief, Ma *et al.* (2013a) directly observed that soot particles could remain open in structure after coating by water. The same particles were compacted after evaporation of the water coating. In the Ma *et al.* (2013a) experiments, restructuring therefore occurred during evaporation, not condensation. Although Ma *et al.* (2013a) did not recommend that their observations should be extrapolated to all atmospheric and combustion systems, many authors have since begun to model coated soot particles as open in structure (Dong *et al.*, 2015; Heinson *et al.*, 2016; Wu *et al.*, 2016, 2017; Wang *et al.*, 2017; Luo *et al.*, 2018; Liu *et al.*, 2016; Lefevre *et al.*, 2019; Luo *et al.*, 2021; Zheng and Wu, 2021; Romshoo *et al.*, 2021). Other authors still model restructuring as occurring upon condensation, in both numerical work (Fierce *et al.*, 2016; Wu *et al.*, 2017; Luo *et al.*, 2018; Kahnert, 2017; Kanngießer and Kahnert, 2018; Luo *et al.*, 2020) and the interpretation of measurements (Cappa *et al.*, 2012; Peng *et al.*, 2016; Zhang *et al.*, 2018; Cappa *et al.*, 2019; Fierce *et al.*, 2020; Yuan *et al.*, 2021). Meanwhile, subtle but clear experimental evidence that condensation does lead to soot restructuring has not been fully acknowledged (Section 3).

The morphology of soot aggregates influences their light absorption and physical properties, as relevant to climate, health impacts and other contexts. Fractal-like soot aggregates scatter up to 30 % more light (Scarnato *et al.*, 2013; Radney *et al.*, 2014), and the optical properties of a coated soot particle which remains open in morphology are significantly different from a compacted one (Scarnato *et al.*, 2013). Experimentally, the relative absorption enhancement caused by coatings on soot is often measured by comparing the absorption properties of coated and denuded soot (Cappa *et al.*, 2012; Lack *et al.*, 2012); this comparison would be biased if the denuded soot was compacted but the coated soot was not. In terms of health impacts, fractal-like aggregates may deposit in the human lung at a different rate than compacted ones (Scheckman and McMurry, 2011; Rissler *et al.*, 2012). The humid environment of the human lungs may potentially trigger compaction for hygroscopic soot particles, influencing their size and surface area, and therefore their health impacts. In a 1975 study, Chamberlain *et al.* (1975) directly observed the compaction of soot aggregates after inhalation and exhalation of engine exhaust by human volunteers. Yet Brauer *et al.* (2001) reported two examples of fractal-like soot found in the lungs of lifetime residents of Mexico City during autopsy. In addition, scrubbers used in exhaust aftertreatment have been observed to cause soot restructuring (Lieke *et al.*, 2013); understanding this behaviour would inform engine aftertreatment system design. Finally, the restructuring of synthetic nanoparticle aggregates is a useful procedure in materials science (Kelesidis *et al.*, 2018).

Overall, there is a strong need to progress towards consensus on the causes and mechanisms of soot aggregate restructuring, as well as to provide a conceptual framework from which the morphological effects of atmospheric aging, cloud processing and human respiration may be predicted. The contribution of the present manuscript towards this goal are to (1) synthesize literature from nucleation, surface and capillary science to describe the processes of phase nucleation, capillary condensation, and droplet formation, (2) critically review experiments on soot aggregate restructuring for evidence of these four modes, (3) present new experiments designed to resolve some of the conflicts in the literature, using solid or liquid anthracene coatings, and liquid oleic acid.

## 2 Review of physical mechanisms required for compaction

Aggregate compaction by condensation must proceed via the (i) nucleation and (ii) growth of a liquid phase, resulting in (iii) the torquing of aggregate monomers due to attractive capillary forces caused by surface tension. These 3 steps have not previously been discussed in the context of the literature on phase nucleation, capillary condensation, and droplet formation. We critically review this literature in the present section. Our review provides a basis for the understanding and classification of the experimental studies on aggregate restructuring reviewed in Section 3.

### 2.1 Nucleation and growth of the liquid phase

Bulk, pure liquid phases may grow when the vapour pressure, $p$, exceeds the equilibrium vapour pressure over a flat surface, $p_0$, at a saturation ratio $S = p/p_0$. Flat liquid surfaces may grow at $S$ slightly greater than 1, while droplets require $S > 1$ due to their positive curvature (Lamb and Verlinde, 2011). In contrast, capillary phases may grow when $S \ll 1$ (e.g. $S < 0.4$; Butt and Kappl 2009; Ferry *et al.* 2002) due to their negative curvature. In all of these cases, the nucleation of the liquid phase is limited by kinetic barriers. These kinetic barriers are important to the process of soot compaction, because they lead to different condensation pathways occurring under different physical conditions. These barriers are described here for the cases of a spherical particle and a capillary pore.

Liquid condensation onto insoluble surfaces is generally initiated by the nucleation of molecular clusters onto active sites on the surface to form small patches of liquid (Laaksonen, 2015; Laaksonen *et al.*, 2020). These small patches have been observed using atomic force microscopy for both high (Cao *et al.*, 2011) and low (Xu *et al.*, 2010) contact angles, and we refer to them collectively as *nanodroplets*.

If the density of nanodroplets formed an insoluble particle has a high enough areal density, they may coalesce to form a film (Laaksonen *et al.*, 2020); the required density is lower for lower contact angles. If a film forms between two nearby surfaces (e.g. of two touching soot spherules), then fluctuations in the film–air interface may lead to the sudden coalescence of a capillary phase (Restagno *et al.*, 2000). The growth of this capillary phase occurs even more slowly than if it were diffusion limited (Kohonen *et al.*, 1999), but may eventually approximate an ideal pendular capillary ring (Christenson, 2001). The nanodroplets, films, and capillary phases may all form at $S < 1$.

At $S > 1$, particles may activate into droplets. This *droplet activation* may proceed via *nanodroplet activation (NDA)*, *film activation*, or *capillary activation* (Laaksonen *et al.*, 2020). Film activation can only occur after film formation, and film formation can only occur when a high density of active sites exist and are given sufficient time for nanodroplets to form and coalesce (Laaksonen, 2015; Laaksonen *et al.*, 2020). Capillary activation can only occur after film formation, and requires that the highly metastable film is given enough time to form a capillary (Restagno *et al.*, 2000). This capillary-formation timescale can be extremely long, as evidenced by the hours required for equilibrium in pore adsorption processes (Restagno *et al.*, 2000; Bocquet *et al.*, 1998). Therefore, metastable supersaturated states are the norm in both capillary nucleation (Restagno *et al.*, 2000; Ferry *et al.*, 2002) and in the adsorption which precedes nucleation (Laaksonen *et al.*, 2020). However, capillary formation rates accelerate exponentially with increasing temperature (Restagno *et al.*, 2000).

Finally, we note that nucleating substances are often observed to be liquid rather than solid, even below the macroscopic melting point. This phenomenon is attributable to the lower free energy of a liquid-solid interface versus a solid-solid interface (Christenson, 2001) as well as the fact that interfacial energy leads to a melting point depression for nanoscopic quantities of matter, as discussed below.

In summary, depending on the nucleation rate and growth rate of the respective phases (which themselves depend on liquid-particle contact angle, density of active sites, and temperature), vapour may condense onto soot particles at $S < 1$ as nanodroplets, films, or capillaries. At $S > 1$, activation of these nanodroplets, films, or capillaries into droplets may result in different shapes of transitory liquid phases. These liquid phases may exert unbalanced surface-tension forces on soot particles, as described in the next subsection.

## 2.2 Mechanisms of soot compaction during condensation

As described in the previous subsection, vapours may condense onto soot particles at $S < 1$ as nanodroplets, films, or capillary condensates. These liquid phases may then activate to form droplets at $S > 1$. During the condensation of these liquid phases at both sub- and supersaturated conditions, the liquid phase may exert surface-tension forces on the soot monomers and cause restructuring of the aggregate. These forces will generally be attractive, and must be asymmetric about the point of contact between spherules for compaction to occur. In this section, we propose specific mechanisms of condensation-induced compaction, focussing on mechanisms for which direct experimental evidence exists in the literature.

The first and second mechanisms are related to capillary formation, and are together referred to as *capillary compaction* in Figure 1. The first occurs at the moment of initial formation of a capillary condensate, when thermal fluctuations in the height of films, adsorbed on surfaces in close proximity, leads to the sudden formation of a capillary connection (Crassous *et al.*, 1994; Restagno *et al.*, 2000). This initial force is about two orders of magnitude larger than the equilibrium capillary force (Crassous *et al.*, 1994) and is therefore not well described by equilibrium-force considerations (Butt and Kappl, 2009), even though such considerations lead to physically reasonable endpoints (Schnitzler *et al.*, 2017).

The second mechanism occurs during condensation at subsaturated conditions. After the liquid phase nucleates, either as a capillary or as a surface film, condensational growth occurs as described previously (Restagno *et al.*, 2000; Crassous *et al.*, 1994; Kohonen *et al.*, 1999). This growth is accompanied by an advancing liquid-solid-gas interface: the contact line. This contact line may advance more quickly in one region of the pendular ring than another, due to physical heterogeneities on the soot surface such as variations in curvature, atomic-scale discontinuities or bumps in the graphitic nanostructure (Toth *et al.*, 2019) or chemically heterogeneous due to surface functional groups (Figueiredo *et al.*, 1999; Matuschek *et al.*, 2007; Corbin *et al.*, 2015b). These heterogeneities mean that the equilibrium contact angle $\theta$ will vary across the soot spherule surface (Gao and McCarthy, 2006), and may also lead to contact line pinning (Fadeev and McCarthy, 1999; Gao and McCarthy, 2006). Pinned contact lines grow via small "jumps" or molecular avalanches (Schäffer and Wong, 2000), which may cause sudden changes and asymmetric torque on the touching spherules. Both pinned contact lines and heterogeneous contact angles would lead to a greater capillary attraction on one side of the pendular ring than the other, resulting in torque on the touching spherules. This second mechanism is dependent on the rate of capillary growth (Eral *et al.*, 2012; Shi *et al.*, 2018) as well as whether or not the touching spherules are already in motion.

The third mechanism is the formation of liquid bridges when pendular capillary rings, or adsorbed films, grow to engulf the particles which they connect. It is labelled in Figure 1 as *bridge-capillary compaction*. This engulfment mechanism was described by Crouzet and Marlow (1995) for a pair of spheres. For an aggregate of spherules, the engulfment would eventually result in contact between the growing capillary ring and three neighbouring spherules. At this point, a similar mechanism to the first may occur. Indirect experimental evidence for this proposition is provided by the two-dimensional analogy observed directly by Narhe and Beysens (2004). However, this mechanism is only weakly constrained by the experimental data, and is included here only for completeness.

During NDA, there is no opportunity for these three mechanisms to act. Therefore, soot particles coated by NDA are unlikely to be compacted (Figure 1), similar to the scenario of coating by coagulation, discussed further in Section 3.

Our first and second proposed mechanisms are consistent with an earlier, apparently intuitive suggestion by Kütz and Schmidt-Ott (1992) which was modelled by Schnitzler *et al.* (2017). While we did not identify evidence in the literature for the formation of a liquid bridge in the shape depicted by Schnitzler *et al.* (2017), we did identify evidence of a mechanism for a mechanism which results in a similar restructuring (third mechanism above) but is limited to immediately neighbouring spherules. Overall, our framework contrasts with the alternative framework proposed by Chen *et al.* (2018), which focussed on the competition between condensational growth rates on surfaces and in capillaries. This competition may come into play in some scenarios, but does not cover all literature cases reviewed below. Although we have proposed these three mechanisms based on directly observed phenomena, the direct observation of these phenomena for nanoparticle aggregates such as soot is not currently achievable experimentally. While our list of mechanisms may not be exhaustive, it is sufficient to describe the substantial body of literature on soot restructuring (Section 3).

## 2.3 Mechanisms of soot compaction during evaporation

The restructuring of a coated aggregate upon evaporation of the coating (evaporative restructuring) occurs due to the attachment of the liquid surface to the aggregate. From the perspective of a single spherule, this scenario is analogous to that of an isolated particle at a liquid–gas interface, for which the theory is well developed (Butt *et al.*, 2003, p. 123). Compared to condensation compaction, the situation is much simpler because it does not involve the nucleation of a new phase.

At equilibrium, such an isolated spherule would be freely floating in the liquid phase due to the balance of buoyant and surface forces. For $\theta > 0$, a small "cap" defined by the contact angle will protrude from the liquid interface (Ref. 229 in Butt *et al.*, 2003). During evaporation, the isolated spherule is normally pulled by the retreating liquid surface. For an aggregate, the critical point of evaporation occurs when two or more spherules experience this pull. This results in a net compacting force on the aggregate, which has been exploited in materials science (Manoharan, 2003; Lauga and Brenner, 2004). For contact angles $> 90°$, this force is still present, the only effect of the high contact angle is that more than half of the spherule area moves to the gas interface (in a bulk liquid, the spherule would appear to float higher). An alternative hypothesis to this description can be made based on the capillary bridges that have been observed between $50\,\mu$m columns in the laboratory (Chen *et al.*, 2017), however, this is much larger than the typical size of agglomerated aerosols.

In the natural world, the most common context for this discussion is the compaction of soot. In one scenario, this may occur when volatile organics emitted from combustion (including internal combustion engines) condense onto soot as the combustion emissions cool. The organics may be non-polar compounds derived from fuel or lubrication oil (Cain *et al.*, 2010) and the soot may be initially hydrophobic. If both the soot surface and the organics are hydrophobic, a low $\theta$ would be expected, and condensation-compaction (Figure 1a and b) may occur. Once emitted to the atmosphere, both the volatile organics (Vakkari *et al.*, 2014) and the soot surface (Corbin *et al.*, 2015b) may become functionalized with polar functional groups, decreasing their hydrophobicity. This atmospheric aging occurs on the timescale of hours (Vakkari *et al.*, 2014). As both gas and particle would be oxidized simultaneously, it is unlikely that $\theta$ between the aged organics and aged soot would remain high, as shown experimentally by (Schnitzler *et al.*, 2014). Therefore, the conditions for condensation without compaction (Figure 1c) are unlikely to occur in fresh or aged combustion systems.

## 3 Review of experimental demonstrations of aggregate restructuring mechanisms

Dozens of studies have either focussed on or commented on the restructuring of soot after exposure to condensable vapours. In this section, we review these studies in light of the four mechanisms identified above and summarized in Figure 1. For completeness, we also include here those relatively few studies which have investigated the restructuring of aggregates or agglomerates other than soot. We find that the literature is consistent with the mechanisms of capillary compaction, bridge compaction, and evaporation compaction shown in Figure 1, with very few exceptions. For completeness, we also mention a fourth mechanism (laser-induced restructuring) and address reported exceptions (no restructuring).

### 3.1 Background: electron microscopy, mobility size, and shape factor

Before reviewing the literature in earnest, we briefly introduce two of the most common techniques for compaction measurements. The first technique, electron microscopy, is widely used, but requires low pressure conditions (vacuum or near vacuum) for analysis, which triggers the evaporation of volatile coatings. The electron beam may also trigger evaporation itself, which hinders the imaging of coated soot aggregates. Additionally, the labour intensity nature of microscopy means that such studies typically report measurements on tens of particles, rather than tens of thousands of particles as in the second common technique.

The second common technique, mobility diameter measurements, measures the migration velocity of particles in an electric field. The resulting $d_{mob}$ is the product of the shape factor $\chi$ and the spherical-equivalent diameter of the particle, after correcting for the non-continuum nature of the gas phase (Equation S1). The shape factor $\chi$ describes the ratio of the drag force on the particle to the drag force on an equivalent-volume sphere. When a particle's material density is known, $\chi$ can be calculated from measurements of particle mass and mobility (Equation S3). The shape factor $\chi = 1$ for spheres (e.g. liquid particles). For a 300 nm soot particle, $\chi$ can be as large as 2 or 3 (Sorensen, 2011, and our Figure 5). Following compaction, $\chi$ is about 1.5 for moderately large soot particles (Figure 5; Ghazi and Olfert 2013), although this limit is smaller for aggregates containing smaller numbers of spherules.

## 3.2 Evidence for capillary compaction

A large number of studies have reported a decrease in the mobility diameter $d_{mob}$ of soot aggregates after the condensation (and sometimes both condensation and evaporation) of liquid coatings (e.g. Kütz and Schmidt-Ott, 1992; Weingartner *et al.*, 1995, 1997; Gysel, 2003; Zhang *et al.*, 2008; Khalizov *et al.*, 2009; Xue *et al.*, 2009; Pagels *et al.*, 2009; Miljevic *et al.*, 2010; Tritscher *et al.*, 2011; Bambha *et al.*, 2013; Ghazi and Olfert, 2013; Peng *et al.*, 2016; Leung *et al.*, 2017b; Pei *et al.*, 2018). Only a few of these studies provide direct evidence for capillary compaction, as distinct from evaporation. These will be discussed here.

### 3.2.1 Use of $d_{mob}$ to infer compaction

The mobility diameter $d_{mob}$ is not a direct measurement of compaction. It may be hypothesized that $d_{mob}$ may change upon condensation due to a change in $\chi$ rather than a change in size. However, Leung *et al.* (2017b) showed unequivocally that a decrease in $d_{mob}$ is due to condensation compaction and is not a measurement artifact. Leung *et al.* (2017b) formed *p*-xylene SOA at extremely low RH ($< 12\%$), where it is known to form a glassy solid (Song *et al.*, 2016). At high RH, the SOA would have been liquid (Song *et al.*, 2016).

Leung *et al.* (2017b) showed that $d_{mob}$ decreased with the addition of small amounts of liquid SOA, but increased with the addition of small amounts of solid SOA. Therefore, the liquid condensation must have caused compaction. This result is consistent with the anthracene results discussed in Section 4, however, we note that solid SOA is glassy (Järvinen *et al.*, 2016) while solid anthracene is crystalline and may therefore cause a faster increase in $d_{mob}$, if both solids were formed by condensation.

It is also important to note that Leung *et al.* (2017b) also observed soot compaction at relative humidities just slightly above 20%. At these low humidities, the coating viscosity is likely 5 orders of magnitude higher than at 80% humidity (Song *et al.*, 2016). This suggests that changes in viscosity are of secondary importance to restructuring phenomena.

### 3.2.2 Condensation below saturation

Miljevic *et al.* (2010) observed that freshly-produced soot particles compacted when bubbled through a variety of organic vapours, but not when bubbled through water. These results are particularly important because of three details of the bubbling technique, which stand in contrast to the more-common alternative of exposing the soot particles to supersaturated vapours.

First, the vapour pressure within a bubble is slightly below saturation, due to the inverse Kelvin effect (relatively small at large diameter of most bubbles). Since compaction occurred below saturation, and since capillary menisci condense well below saturation, this corroborates the capillary-condensation mechanism described above and suggests a role of contact angle (which governs capillary formation).

Second, when Miljevic *et al.* (2010) switched from bubbling candle soot through hexane to bubbling through water, the soot did not undergo compaction. This observation is consistent with the Ma *et al.* (2013a) study discussed below, and illustrates the key role of contact angle. The high water-soot contact angle would have prevented capillary condensation from occurring. Third, condensation occurred at room temperature in 1 second, suggesting that metastable effects were negligible for their soot sample.

In a separate pair of studies, Chen et al. (Chen *et al.*, 2016, 2018) showed that a sub-monolayer volume of PAHs coated onto soot could cause substantial compaction. As this sub-monolayer could only have exerted capillary compaction forces (Section 2), these experiments provide clear evidence for that mechanism, as also concluded by those authors.

### 3.3 Evidence for bridge compaction

Chen *et al.* (2018) presented an important data set which clearly illustrates the difference between capillary (Fig,. 1a) and bridge-capillary (Fig,. 1b) compaction. Whereas a first group of compounds (Group A) caused compaction (reduced $d_{mob}$) with minute coating volumes ($\sim 5\%$), a second group (Group B) required much larger coating volumes (10–48%) to be compacted. In Group A, compactness increased smoothly with coating volume, up to double the initial particle volume. In Group B, compactness increased suddenly with small coating volumes, and the addition of greater coating volumes did not cause further compaction. Whereas the compactness of Group B compounds increased smoothly with increasing coating volumes, the compactness of Group A compounds reached a minimum rapidly. Chen *et al.* (2018) noted that surface tension differences could not explain these trends, consistent with our inference from Leung *et al.* (2017b) above. We also note that Groups A and B contained compounds of a range of polarity, which implies a range of contact angles.

The one essential difference between Groups A and B in Chen *et al.* (2018) is that the Group A compounds had vapour pressures ranging from $0.18$ to $2.00\,\mathrm{Pa}$, whereas Group B compounds had vapour pressures ranging from $10^{-7}$ to $10^{-4}\,\mathrm{Pa}$ at $25\,^\circ C$. Consequently, Chen *et al.* (2018) exposed soot particles to Group B compounds at much higher experimental temperatures ($\geq 50\,^\circ\mathrm{C}$) compared with Group A. This difference in temperature altered the coating volume, as pointed out by Chen *et al.* (2018), but also created much higher transient saturation ratios. According to the calculations of Chen *et al.* (2018), the Group-B temperatures resulted in maximum saturation ratios of $S_{max} > 20$ with less than 1 second spent between $S = 0.1$ and $S = 1$ prior to coating formation. Consequently, condensation may have occurred too quickly for capillary nucleation to occur. This suggests that Group A and Group B compounds triggered soot restructuring through capillary compaction and capillary-bridge compaction, respectively (cf. Figure 1).

In their discussion and subsequent work (Ivanova *et al.*, 2020), Chen *et al.* (2018) focussed on the competition between the Kelvin effect and vapor supersaturation. As their data set also varied $S_{max}$ substantially, it is not clear whether this competition played a key role. Additional data to constrain this hypothesis would be valuable. We note that although atmospheric SOA may also reach high supersaturations ($S > 4$ for the lowest-volatility compounds, Donahue *et al.* 2011), environmental chamber data on soot compaction have already provided clear evidence that condensation-compaction occurs in the atmosphere (Qiu *et al.*, 2012; Schnitzler *et al.*, 2014; Guo *et al.*, 2016; Leung *et al.*, 2017b). The supersaturations used in such laboratory studies are therefore likely to be representative of atmospheric conditions.

### 3.4 Evidence for evaporation compaction and NDAC

Three studies have indicated evidence for evaporation compaction. Two of these data sets demonstrate only subtle evidence, while the third presents direct evidence. The subtler evidence from Leung *et al.* (2017b) and Chen *et al.* (2016), using glassy SOA and anthracene respectively, showed that $d_{mob}$ increased for thin coatings but decreased when these coatings were denuded, most likely due to melting in the denuder. The glassy SOA case was discussed above; the anthracene result is returned to below in Section 4.3, and Figures 2 and 6.

The third study provides the most direct evidence, by observing the morphology of water-coated soot by injecting it into bulk water and measuring the particle fractal dimension by static light scattering (Ma *et al.*, 2013a). Ma *et al.* (2013a) found that this fractal dimension remained low within the water droplets (less than 1.94), corresponding to an open morphology. They also demonstrated that the same soot particles underwent compaction during water evaporation using transmission electron microscopy (TEM) and by measuring a mass-mobility exponent of 2.79 (see Sorensen 2011 and Olfert and Rogak 2019 for a detailed discussion of the mass-mobility exponent).

In the context of Section 2, it is extremely likely that Ma *et al.* (2013a) observed NDAC (Figure 1), which does not cause compaction. When evaporating the water, they observed evaporation compaction. The soot particles studied by Ma *et al.* (2013a) did not take up water below $S = 1.2$. This shows that they were hydrophobic. Hydrophobicity implies a high contact angle ($\theta > 90°$) as expected from literature (Persiantseva *et al.*, 2004). Therefore, capillary condensation would not have occurred, and film formation would be unlikely. Rather, water will have condensed via NDAC (Figure 1), as described above.

There are two reasons why the results of Ma *et al.* (2013a) should not be extrapolated to atmospheric science. First, such high $S$ do not normally occur in the atmosphere, where most particles activate at $S < 1.01$ (Schmale *et al.*, 2018). Second, atmospheric soot typically becomes rapidly oxidized or coated with small amounts of hydrophilic material (e.g. Vakkari *et al.*, 2014) which are observed to result in water condensation water at much lower $S$ (Mikhailov *et al.*, 2006; Xue *et al.*, 2009). Surface oxidation may also occur (Matuschek *et al.*, 2007; Corbin *et al.*, 2015a). These hydrophilic material or oxidized groups may provide nucleation sites for film nucleation to occur (Section 2). Atmospherically realistic condensates like secondary organic aerosol have been shown to condense readily onto soot, without requiring extreme supersaturations (Qiu *et al.*, 2012; Schnitzler *et al.*, 2014; Guo *et al.*, 2016; Leung *et al.*, 2017b), partly because of the presence of organics with a broad range of volatilities (Tröstl *et al.*, 2016). Thus, the work of Ma *et al.* (2013a) represents a specialized system in which the phenomenon of evaporation-compaction could be demonstrated.

We note that Ma *et al.* (2013a) attempted to explore the role of $\theta$ on their results by oxidizing their soot samples in a furnace. They oxidized soot particles at 300, 600, and 700 °C. However, as shown in a later study using the same system (Ma *et al.*, 2013b), at these temperatures soot oxidation results in a decrease of particle mass due to volume oxidation by $O_2$, in contrast to the surface oxidation that occurs at higher temperatures (Kelesidis and Pratsinis, 2019) . This decrease of particle mass is evident in the oxidation results of Ma *et al.* (2013a) and makes the quantitative interpretation of their data is difficult. Nevertheless, it is clear that this furnace treatment did not have the expected result of rendering the soot particles hydrophilic, since the maximum saturation ratio required to activate particles after furnace oxidation was even higher ($S > 1.5$) after furnace oxidation compared to before ($S > 1.2$). We also note that, at 300 °C, significantly less compaction was observed than for the untreated or 600 or 700 °C cases. This may indicate that surface-bound, semi-volatile, and hydrophobic flame-generated organics such as polyaromatic hydrocarbons (Slowik *et al.*, 2004; Cain *et al.*, 2010) played a role in the results of (Ma *et al.*, 2013a), for example, by acting as active sites for nanodroplet activation.

### 3.5   Other notable studies

Olfert and Rogak (2019) reviewed a large number of studies and showed that the effective density trends with $d_{mob}$ are similar for many engines and laboratory flames. However, they showed that the effective density (a proxy for compactness) of compression ignition engines was generally slightly higher, especially for small particles. This suggests that some degree of compaction occurs in those engines, since condensation is most rapid for smaller particles. Further research should explore this possibility.

### 3.6   Other restructuring phenomena: laser-induced decompaction and sintering

Though unrelated to compaction by coatings, here we briefly mention two other restructuring phenomena. First, Bambha *et al.* (2013) reported that soot restructuring could be reversed by laser-heating. They coated soot by condensation, then used a pulsed 1064 nm laser to heat the soot and cause coating evaporation. Since condensation compaction is to be expected, this implies that the soot particles underwent *decompaction* due to the laser heating process, perhaps due to the nucleation of bubbles at the soot surface (in analogy to heterogeneous nucleation during boiling). Such bubbles would remain attached to the soot surface in order to minimize their Gibbs energy, and may lead to expansion of the soot structure during their growth.

Bambha *et al.* (2013) also stated that they observed compacted soot in the electron microscope after coating, but they were unable to image the particles before coatings evaporated. Such images would constitute evidence for condensational compaction, if they could be produced without any interference of the evaporating droplet.

Second, aggregates of metal nanoparticles are known to sinter (partially coalesce without liquefying) at temperatures well below the melting point, and this leads to aggregate compaction (Kleinwechter *et al.*, 1997; Schmidt-Ott, 1988). The same is true for metal-oxide nanoparticles (Kelesidis *et al.*, 2018). This process is attributable to melting-point depression (Equation S4) and not capillary forces. Conversely, the strong inter-particle bonds of sintered aggregates can prevent compaction (Kelesidis *et al.*, 2018).

Finally, we briefly mention the study of Weingartner *et al.* (1997) who observed the compaction of spark-generated carbon nanoparticles at $S < 1$. These spark-generated nanoparticles have significantly different properties to soot (Gysel *et al.*, 2012) so are not discussed in the soot section above.

## 3.7 Avoiding compaction: observations of exceptions

A small number of studies appear to imply that compaction can be avoided in some cases. All of these studies have used microscopy, rather than in-situ methods, and have therefore been limited to extremely small numbers of particles. Caution should therefore be exercised when interpreting these studies as generally representative.

A salient example of avoided compaction is provided by Cross *et al.* (2010). This study presented a single electron micrograph of a laboratory soot particle coated by dioctyl sebacate and then denuded. The authors suggested that this processing had failed to compact the soot. However, multiple studies using the in-situ $d_{mob}$ technique, which measures orders-of-magnitude more particles, have consistently reported compaction of soot by a similar compound, oleic acid (Ghazi and Olfert, 2013; Chen *et al.*, 2018, and this work). We suggest that the single particle measured by Cross *et al.* (2010) was not representative of the overall particle ensemble. This example is particularly important because it was possible to repeat the measurement, unlike the atmospheric studies mentioned below.

To our knowledge, all other exceptions are represented by very poor sampling statistics (in most cases, between one and three particles in total). These include electron (Huang *et al.*, 1994; Ebert *et al.*, 2002; Mikhailov *et al.*, 2006; Mikhailov and Vlasenko, 2007), X-ray (Zelenay *et al.*, 2011), and atomic force (Köllensperger *et al.*, 1999) microscopy. Since such microscopy typically requires high vacuum conditions, coating evaporation has the opportunity to act prior to sample observation. "Environmental" microscopy has also been used, which uses low pressure rather than high vacuum (Ebert *et al.*, 2002; Zuberi *et al.*, 2005; Zelenay *et al.*, 2011); this technique still cannot observe soot clearly while it is coated.

Microscopy studies have reported observations of internally-mixed but not compacted soot particles (Adachi *et al.*, 2010; China *et al.*, 2013) which may have formed by coagulation. The statistical representativeness of such observations has not been demonstrated. Other examples of obvious coagulation (between soot and dust) have been observed with evidence of compaction (Xiong and Friedlander, 2001). In this scenario, it is possible that condensation onto the soot occurred prior to its coagulation. It is also possible that the soot particle could become encapsulated following coagulation, and thereby have avoided compaction until the coating was evaporated in the microscope. In the future, statistically robust measurements of such morphologies would be valuable.

Here, the study of Brauer *et al.* (2001) deserves special mention. Brauer *et al.* (2001) observed non-compacted soot in the lungs of lifelong Mexico City residents. In the context of Section 2, this observation suggests that these particles entered the lungs as fresh hydrophobic soot, did not nucleate water during inhalation, and deposited onto the respiratory system. In contrast, Chamberlain *et al.* (1975) observed compact soot particles in the exhaled breath of volunteers who inhaled the exhaust of an older gasoline engine burning leaded fuel. The resulting particles might be more toxic, due to their increased active surface area, and may interact differently with human macrophages. These observations can be interpreted as suggesting that coagulation also does not result in compaction.

The fraction of coagulated particles in the atmosphere is therefore a reasonable first-order estimate for the fraction of non-compact-but-coated soot. Coagulation is an important mixing mechanism for soot during photochemically inactive periods (e.g. at night) (Fierce *et al.*, 2015) as well as in heavily polluted regions such as Eastern China, India, and Central Africa (He *et al.*, 2016). After coagulation, subsequent coating growth (including hygroscopic growth) may lead to encapsulation and eventual compaction of the soot particle. Future studies should address these processes, and assess to what degree they may affect soot light absorption.

## 4   Results

In this Section we present the results of a series of experiments designed to separately study condensation and evaporative compaction. The key element of these experiments was the "switching off" of surface tension by selectively freezing and melting a surrogate coating material before evaporation or sublimation. As discussed below, the results demonstrate that both condensation and evaporation can cause soot restructuring.

### 4.1   Experimental design

#### 4.1.1   Compaction pathways

The goal of the experiments presented below was to isolate the effects of surface tension on soot compaction during coating addition versus coating removal. The experiments are summarized in Table 1, which introduces the labels used in the subsequent discussion. These four experimental pathways are depicted using the phase diagram in Figure 2 and in a simple schematic in Figure 3. The lines on this phase diagram were calculated as described in Section S1.2.

The four pathways in Table 1 represent combinations of liquid condensation or evaporation (**L**) and solid deposition or sublimation (**S**). These four steps (adding and removing **S** or **L**) are shown on Figure 2 using five curves, since there are two options for liquid condensation (capillary condensation or NDA; Section 2). Pathways **LS** and **SL** isolate the effects of surface tension to only the condensation or evaporation processes, respectively. For Pathway **LS**, we performed experiments where liquid coatings were condensed, then frozen prior to their removal as solids (sublimation). Since surface tension cannot act during sublimation, we attribute any restructuring via Path **LS** to the condensation process. Path **SL** is the inverse of Path **LS**. Solid coatings are deposited, but are melted prior to their removal as liquids (*evaporation*). So, surface tension acts only during coating removal. Path **SS** is a control experiment where no surface tension forces are allowed to act. Coatings are added in the solid phase (*deposition*) and subsequently sublimated. Here, restructuring should not occur, as surface tension forces never come into play. The use of the phase diagram in Figure 2 to design our experiments is further discussed in Section S1.2.1.

#### 4.1.2   Experimental methods

Our experimental setup (Figure S1) was similar to that used in previous studies (e.g. Nguyen *et al.*, 1987; Moteki and Kondo, 2007; Slowik *et al.*, 2007; Chen *et al.*, 2016). The present subsection therefore summarizes only the key experimental details of our experiments; complete details are given in the supplementary information (Section S1.3).

Soot was generated by a miniCAST 5201c soot generator (Jing Ltd., Switzerland) which consists of a partially-quenched propane diffusion flame. We pre-selected aggregates of mature soot with mobility diameter of 300 nm before using coating apparatuses constructed in-house to add anthracene or oleic acid coatings to the soot. The coating apparatuses generally consisted of a heated section (using either a hot plate, oil immersion, or heating tap) followed by a cooling section (using either heat conduction to room air, additional insulation to slow cooling, or ice packs to accelerate cooling). Oleic acid coatings were added with the apparatus heater set between 94 and 140 °C. Anthracene coatings were added with the apparatus heater set between 95 and 202 °C. Above 202 °C, the reservoir emptied too quickly for measurements to be practical. In some experiments, we pre-coated the walls of the second apparatus with anthracene in order to increase its vapour phase concentrations, which allowed us to avoid anthracene sublimation upon heating and to melt anthracene prior to its evaporation.

The primary denuder used in this work was a catalytic stripper (CS015, Catalytic Instruments GmbH, Rosenheim, Germany), which vapourizes and then oxidizes organic molecules at 350 °C. In some experiments, this stripper was replaced with the activated-charcoal thermodenuder described by Burtscher *et al.* (2001) or a second coating apparatus, as described in Section S1.3.

## 4.2 Observation of condensation and evaporative compaction

### 4.2.1 Size distribution changes

Figure 4 shows $d_{mob}$ distributions measured during the experiments described in Section 4.1. The experiments used anthracene or oleic acid to explore the four pathways of Table 1 and Figure 3. Oleic acid was used for Path **LL**; anthracene was used for all others. The inset TEM images in Figure 4 depict aggregates sampled from each condition. Not all particles observed in the TEM were fully compacted after processing (Figure S3); this may be due to variability between compaction events or heterogeneity in the conditions of our processing apparatus. We consider those particles as outliers considering that the size distributions include orders of magnitude more particles and that the TEM analysis is subject to operator bias. Figure 4a also includes an example unprocessed particle.

Figure 4c intentionally presents conditions where a very large volume of solid coatings was added. For consistency, the comparable Figure 4d uses similarly large volumes. These large volumes were selected in order to ensure that the lack of observed compaction was not due to a lack of sufficient coating material. In contrast, Figure 4a and Figure 4b show relatively thin coatings, as these were sufficient to induce compaction. The coating volumes are indicated by the text above each size distribution, which show that the "large" and "relatively thin" coatings just mentioned were 2-fold and 10-fold the original particle volume, respectively, as measured by single-particle mass measurements. Compaction trends with coating volume are discussed in the next section.

Figure 4c shows results for the addition and removal of solid coatings via Path **SS** of Table 1 and Figure 3. The results show that Path **SS**, coating deposition–sublimation, resulted in no change to $d_{mob}$. The lack of compaction for Path **SS** shows that surface tension forces are necessary for compaction in general, and rules out the hypothesis that van der Waal's forces alone are sufficient.

In Figure 4d, surface-tension-related forces were allowed to act since coatings were melted prior to their evaporation (Path **SL**). This resulted in a reduced final $d_{mob}$ and compacted particles, as confirmed by TEM.

In Figure 4b, the reverse of Figure 4d was measured. Liquid coatings were added by condensation, but frozen prior to their removal by sublimation (Path **LS**). Again, this resulted in a reduced final $d_{mob}$ and compacted particles, as confirmed by TEM. Since capillary forces cannot act during sublimation, this confirms that condensation compaction occurred in Path **LS**. Thus, our experiments demonstrate that both condensation compaction and evaporative compaction may occur: aggregate compaction may be triggered by liquid condensation or liquid evaporation, but not solid deposition nor solid sublimation. This conclusion is corroborated by previous studies using anthracene in Section 4.3.

Finally, Figure 4a shows results for Path **LL** (coating condensation and evaporation). Compacted particles were observed after this experiment. Most of this compaction likely occurred during condensation, simply because condensation occurred before evaporation.

### 4.2.2 Coating volume effects

Figure 5a shows the results of Figure 4 in terms of the change in modal $d_{mob}$ as a function of coating volume. That is, Figure 5a extends Figure 4 to a much larger number of comparable experiments in which the coating mass was varied. Data are shown for both anthracene and oleic acid coatings. Each group of data is labelled with two letters following Table 1: the first letter (S or L) indicates the coating phase during addition, the second letter (S or L) indicates the coating phase during removal. Curves are included to guide the eye. Error bars are included for sparse data and are omitted where multiple similar measurements provide an indication of measurement precision.

In Figure 5, the mode $d_{mob}$ was retrieved from fits to the data. When more than one mode was apparent due to heterogeneity in coating thicknesses (e.g. Figure 4, panel 3 and 4), the mode containing more particles was reported, for consistency with the denuded size distributions. Figure 5b plots the same data after conversion to shape factor $\chi$ using Equation S1.

Figure 5a shows that oleic acid condensation led to a decrease of $d_{mob}$ until the particle volume approximately doubled ($V_1/V_0 > 2$), that is, until the particle contained equal volumes of coating and soot. For thin coatings, $d_{mob}$ was similar for coated and denuded particles. For these particles, the shape factor $\chi$ (Figure 5b) was larger for denuded than for coated particles. In the context of Sections 2 and 3, this is interpreted as condensation compaction; specifically, the BCC mechanism (Figure 1b). The alternative, the CC mechanism (Figure 1a) would have occurred for much thinner coatings and would appear as a much steeper slope in the figure.

Figure 5b shows that $\chi$ decreased monotonically with increasing $V_1/V_0$ for both coating materials. This $\chi$ reached an asymptotic value of unity, demonstrating particle sphericity for oleic acid coatings. Sphericity is expected when the particle surface is liquid, and when aggregates are fully encapsulated. Importantly, $\chi = 1$ was also reached for liquid-anthracene coatings. This was not observed for solid anthracene coatings, where $\chi = 1.2$ was the asymptotic value. Slowik *et al.* (2007) achieved a similar asymptotic value ($\chi = 1.3$) for solid anthracene (Figure S5). Therefore, our measured shape factors support our conclusion that we were able to deposit solid or liquid anthracene coatings by varying our experimental conditions. For all $V_1/V_0$, $\chi$ is higher for denuded than for coated particles because a compact soot aggregate is not perfectly spherical (Figure S3).

Because our coating apparatus did not allow us to finely control the amount of anthracene deposition in experiments related to Paths **3** and **4**, only a single data point is available for those cases. Our limited data for Paths **3** and **4** is nevertheless corroborated by the literature comparison below.

The main difference of our work with previous literature is that we observed smaller amounts of restructuring for the removal of solid coatings than liquid coatings. However, because we were only able to follow Path **1** using oleic acid, and not anthracene, we cannot determine whether this small difference is due to surface tension effects (Schnitzler *et al.*, 2017), contact angle effects, or differences in the restructuring mechanism.

### 4.2.3 Further validation experiments

To confirm that bulk anthracene sublimated upon heating in our apparatus, we performed an experiment where anthracene powder (as received) was heated in the coating-apparatus reservoir while the apparatus was kept open (in a fume hood) for observation. The crystals were heated continuously in air at roughly $1\,°C\,s^{-1}$. We observed no visual changes until the apparatus reached $191\,°C$, at which point the crystals began to shrink rapidly. As the temperature was increased to $220\,°C$, fumes were observed without any indication of melting. A spatula placed within the fumes became coated with a matte film initially and millimetre-sized planar crystals later (Figure S4). The initial matte film on the spatula may be related to the melting point reduction of nanoscopic quantities of anthracene (Figure S2). Although our visual observation could not confirm this hypothesis, Lopatkin *et al.* (1977) have directly observed that thin films of anthracene condense as as liquids at temperatures above $55\,°C$. Lopatkin *et al.* (1977) studied films of thickness $10\,nm$ to $16\,\mu m$ with growth rates of 3 to 15 nm/s. Under these conditions, Lopatkin *et al.* (1977) reported that these anthracene films deposited as solids below $55\,°C$. This is consistent with our melting-point depression calculations, discussed above and in the SI.

To ensure that the phase of condensing anthracene was not affected by residual OM from the flame, we pre-denuded the soot at $350\,°C$ before repeating the condensation ($V_1/V_0 \sim 3.3$) and condensation–denuding (at $200\,°C$ and $150\,°C$) experiments. No change was observed in our results.

We also ensured that anthracene oxidation was not relevant in our apparatus by repeating the experiment in nitrogen rather than synthetic air. The results were included in Figure 5 at $V_1/V_0 = 2.55$ and are fully consistent with the other results.

### 4.3 Comparison with previous soot–anthracene studies

Two previous studies have reported data similar to Figure 5 for anthracene (Slowik *et al.*, 2007; Chen *et al.*, 2016). These data have been included in Figure 6. The purple diamonds in Figure 6 show the data of Slowik *et al.* (2007), who coated soot with anthracene at 52–85 °C. As shown in Figure S2, this is well below the melting point of even 3 nm anthracene spheres, so Slowik *et al.* (2007) most likely deposited solid coatings. Slowik *et al.* (2007) removed these coatings at 200 °C, which is below the 216 °C melting point of anthracene. Therefore, as also concluded by those authors, the anthracene coatings of Slowik *et al.* (2007) did not result in restructuring of soot; they followed Path **SS** of Figure 3.

Chen *et al.* (2016) coated soot with anthracene (and 5 other PAHs) at 25–85 °C. They denuded their soot at 300 °C, which is well above the 216 °C melting point of anthracene. Their coated soot increased in $d_{mob}$ substantially, and their denuded soot decreased in diameter substantially, consistent with our results and those of Slowik *et al.* (2007). Therefore, as also concluded by those authors, Chen *et al.* (2016) did not restructure soot by anthracene deposition. Although not discussed by those authors, anthracene evaporation did restructure soot. Therefore, Chen *et al.* (2016) followed Path **4** of Figure 3.

We have not compared our results with Slowik *et al.* (2007) or Chen *et al.* (2016) in terms of shape factor $\chi$, because those two studies and our study used soot with different spherule diameters and aggregate mobility diameters. So, the initial and final $\chi$ expected in the three studies is different, as shown systematically by Leung *et al.* (2017a). Regardless, we have reanalyzed the data of Slowik *et al.* (2007) in Figure S5 to show that they do illustrate the same trends observed here. For the data of Chen *et al.* (2016), shape factor changes are small since those authors focussed on thin coatings.

Here it is worth repeating the statement by Chen *et al.* (2016) that complex mixtures are likely to have lower melting points (Peters *et al.*, 1997; Marcolli *et al.*, 2004). Nanoscopic quantities of material are also more likely to exist in the liquid phase (Section S1.4). Consequently, in combustion and atmospheric chemistry, coatings are most likely to condense as liquids. Important exceptions may be the scenarios where secondary organic aerosols reach very high viscosities due to low temperatures or humidity (Leung *et al.*, 2017b; Koop *et al.*, 2011; Schmedding *et al.*, 2020) and where ice crystals may form without transitioning through cloud droplets on soot (Marcolli, 2014; David *et al.*, 2019).

## 5 Summary

An understanding of soot compaction based on well-established, directly observed phenomena in surface science (Section 2) indicates that there are two mechanisms for condensation compaction (capillary and capillary-bridge) which regularly occur in the environment. A third mechanism, nanodroplet activation, requires extreme supersaturations and has been demonstrated in the laboratory, but not observed in natural systems. If condensation compaction is avoided (by the third mechanism or by coagulation), evaporation also leads to compaction.

This framework of condensation-compaction and evaporation-compaction mechanisms resolves conflicting conclusions in the literature (Section 3) and was demonstrated in the laboratory (Section 4). Future modelling studies focussed on atmospheric and climate science should assume that soot particles become compact when coated.

*Author contributions.* JCC conceived the study, initiated the experiments and critical reviews, and drafted the paper. RLM and MGB discussed initial results, co-designed subsequent experiments, and contributed substantially to data interpretation and to writing.

*Acknowledgements.* This work was funded by the ERC under grant ERC-CoG-615922-BLACARAT. Thanks are owed to Maarten Heringa for providing a component of the condensation apparatus, to Elisabeth Müller for TEM assistance, and to Louis Tiefenauer for the loan of the TEM sampler. We are grateful to Jay Slowik and Alexei Khalizov for their openness in sharing published data and to Ogochukwu Y. Enekwizu for stimulating discussions.

**Table 1.** Summary of experimental pathways towards aggregate compaction. Literature evidence for each pathway is discussed in Section 3. This table summarizes how the mechanisms in Figure 1 were used with the phase diagram in Figure 2 to achieve the experimental pathways shown in Figure 3. The contents of this table are depicted graphically in Figure 3.

| $\theta$ [°][b] | Coating added as | Coating removed as | Compaction mechanisms[a] | Material used in this study |
|---|---|---|---|---|
| n.a.[b] | Solid | Solid | — | Anthracene |
| < 90 | Solid | Liquid | DE | Anthracene |
| < 90 | Liquid | Solid | CC, BCC | Anthracene |
| < 90 | Liquid | Liquid | CC, BCC, and/or DEC | Oleic acid |
| > 90 | Liquid | Liquid | NDAC | n.a. |

[a] Acronyms for each compaction mechanism are defined in Figure 1. [b] n.a. = not applicable.

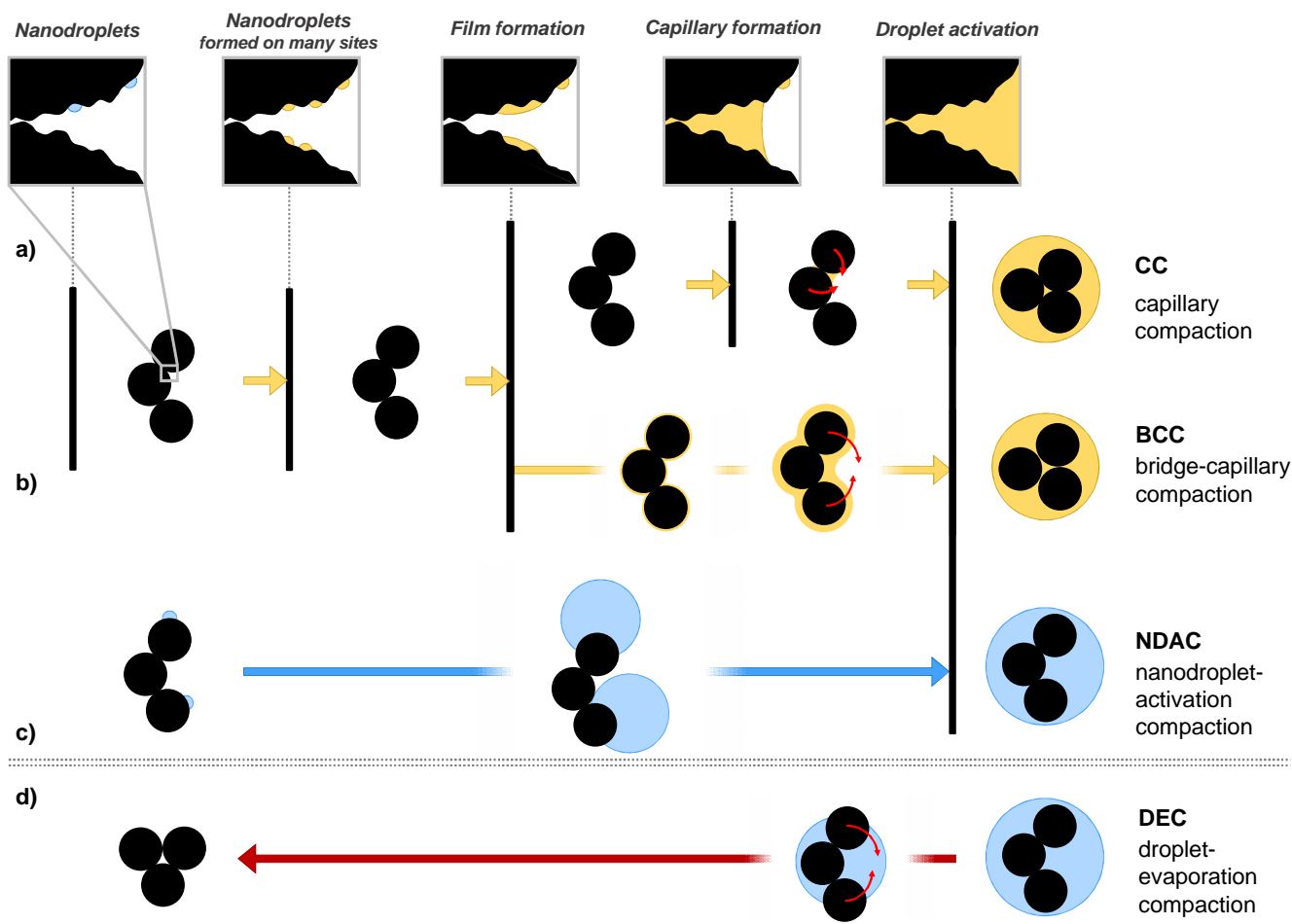

**Figure 1.** Compaction mechanisms for soot in the context of the activation barriers (top row, italicized labels) involved in forming the liquid phase. These mechanisms are based on the experimental observations reviewed in Section 2 and are generally sensitive to both time (kinetic limitations) and vapour saturation (thermodynamic limitations). Yellow and blue colours indicate liquids with low and high contact angles, respectively. Dark-red arrows indicate evaporation, which is only depicted for the case where compaction was avoided during condensation. Light-red curved arrows show torque due to compaction forces. In the top row, surface heterogeneities that may act as adsorption sites are depicted by the jagged line. For NDAC, the critical droplet diameter prior to activation is drawn at a scale that represents an approximate critical diameter, assuming a 30 nm diameter soot spherule.

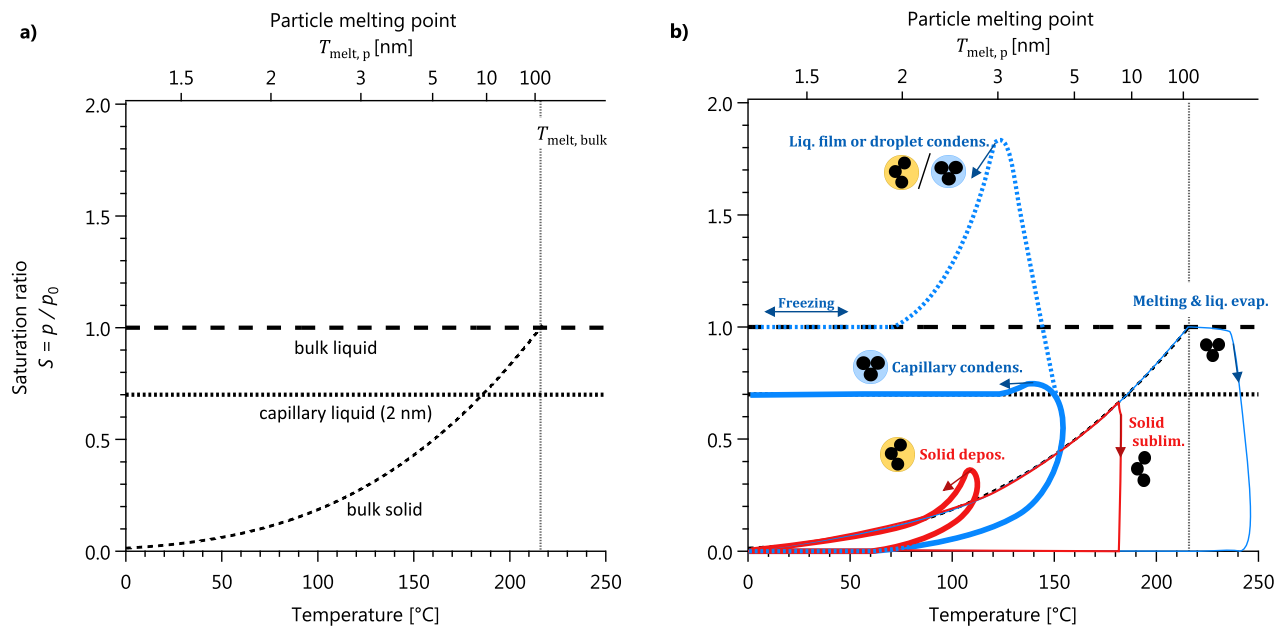

**Figure 2.** Equilibrium phase diagram for anthracene, as used to design the experiments shown in Figure 3. Saturation ratios are plotted with respect to the bulk liquid vapour pressure, $p_0$. Panel **a** shows the equilibrium saturation ratios $S$ for bulk solids, bulk liquids, capillary liquids with $> 2\,\text{nm}$ equivalent diameters and the melting-point depression associated with nanoscopic quantities of material (Section S1.2.1). Nanoscopic quantities of material are expressed in equivalent nanoparticle diameters on the top axes. Panel **b** qualitatively illustrates the experimental pathways used to add (thick lines) or remove (thin lines) coatings. Coatings are either solid (red) or liquid (blue) phase at the time of removal. By combining one addition and one removal pathway, different compaction endpoints can be achieved, as shown by the particles. Note that the depicted maximum supersaturations are conservative; some laboratory studies have used $S > 20$.

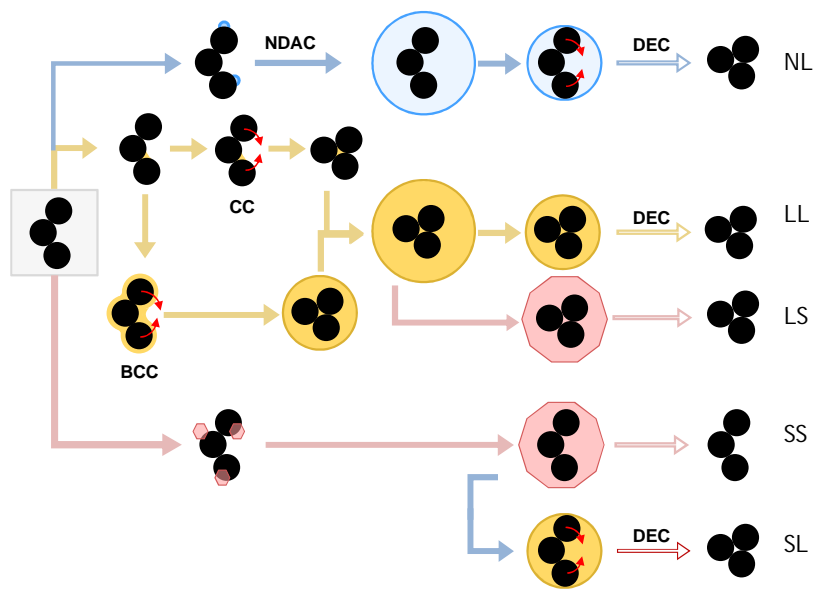

**Figure 3.** Experimental pathways towards soot compaction by condensation or evaporation. This figure combines the concepts in Table 1 and Figure 1. Blue and red arrows indicate liquid and solid phase coatings, respectively. Solid arrows indicate coating addition, open arrows coating removal. Red, blue, and orange coatings indicate solid, low-$\theta$ liquid, and high-$\theta$ liquid phases, respectively. Red text indicates opportunities for compaction (mechanisms shown in Figure 1) and black text indicates observed experimental outcomes (Figure 5). In our experiments, nanodroplet-LL was not achieved due to the low contact angles between our coating materials (oleic acid and anthracene) with soot.

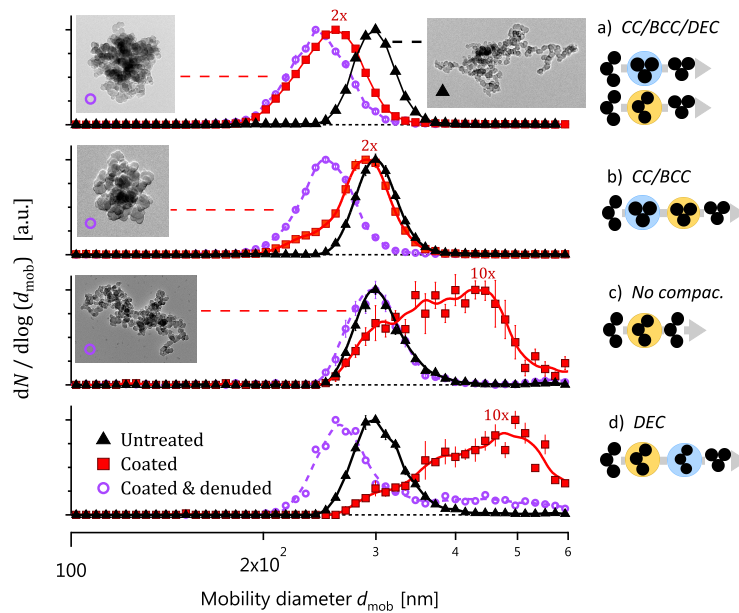

**Figure 4.** Size distributions of 300 nm soot particles without treatment (▲), after addition of solid or liquid organic coatings (■), and after removal of coatings (○). The four panels correspond to different coating phases on addition or removal, which allows different compaction mechanisms (Table 1 and Figure 1) to proceed. Panels **b** and **d**, illustrate compaction by capillary forces (either CC or BCC) and by droplet evaporation (DEC), respectively. Panel (a) was achieved with oleic acid, all others used anthracene. Some curves are labelled with exemplary TEM images (soot monomer diameters $30 \pm 6$ nm) of coated–denuded or untreated samples. The numbers above the coated data points give the volume growth factors (2x, 10x) for comparison with Figure 5. The mechanisms CC and BCC are indistinguishable here, but can be distinguished in Figure 5.

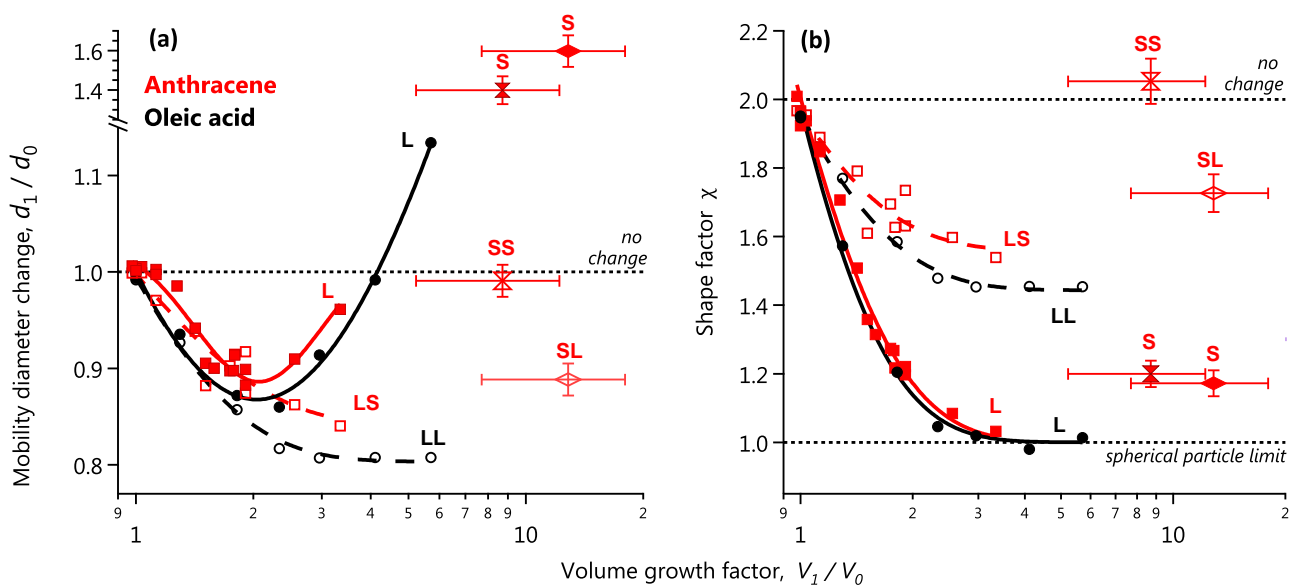

**Figure 5.** **(a)** Mobility diameter $d_{mob}$ and **(b)** shape factor $\chi$ of 300 nm soot as a function of coating volume. Filled symbols show soot particles with coatings added, open symbols show soot particles with coatings added and removed. The coating phase state during addition is indicated by the first (or single) capital letters **S** for solids and **L** for liquids. The coating phase during removal is indicated by the second capital letters. Relative to Figure 4, this additional data suggests that liquid coating addition (**L**) resulted in BCC rather than CC, because of the slow decrease in $d_{mob}$ with $V_1/V_0$ (Section 3). In addition, the data suggest that DEC results in more complete compaction than BCC (compare **LL** and **LS**) and that **L** and **S** coatings were thick enough to encompass the entire soot particle (Panel **b**) and reach the spherical-particle limit of 1 in Panel **b**. Fewer data points were available for solid coatings due to the experimental difficulty of adding solid coatings.

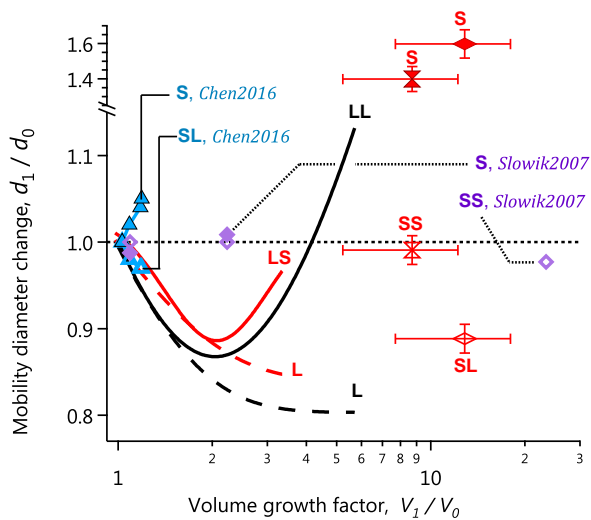

**Figure 6.** Same as Figure 5a but including data of Chen *et al.* (2016) (blue triangles) and Slowik *et al.* (2007) (purple diamonds), which are consistent with Path **4** and Path **3** of Figure 3, respectively. The coated data of Chen *et al.* (2016) indicate solid deposition, since mobility diameters increased upon coating. These three data setse are not plotted here in terms of shape factor $\chi$ (as in Figure 4b) because soot monomer and aggregate sizes differed between studies.

Supporting information for

# "Experimental compaction of soot aggregates by coating condensation and evaporation"

**by Joel C. Corbin, Robin L. Modini, and Martin Gysel-Beer.**

## S1 Supplementary Text

### S1.1 Mobility diameter and particle volume

The mobility diameter $d_{\text{mob}}$ is given by (DeCarlo *et al.*, 2004),

$$\frac{d_{\text{mob}}}{C_c(\text{Kn}, d_{\text{mob}})} = \chi(\text{Kn}) \frac{d_e}{C_c(\text{Kn}, d_e)} \tag{S1}$$

where the dynamic shape factor $\chi(\text{Kn})$ represents the ratio of the drag force on the particle to the drag force on its volume-equivalent sphere with diameter $d_e$, which is commonly obtained from single-particle mass ($m_p$) measurements via $m_p = \rho_p d_e^3 (\pi/6)$ when the particle material density $\rho_p$ is known (DeCarlo *et al.*, 2004). The function $C_c(\text{Kn}, d)$ is the Cunningham slip correction, which accounts for the non-continuous nature of the gas-phase and which depends on the Knudsen number. $\chi(\text{Kn})$ is a function of the Knudsen number (Kn $= 2\lambda/d$, where $\lambda$ is the molecular mean free path of air and $d$ is a particle diameter). For simplicity, we generally write $\chi$.

### S1.2 Calculation of the phase diagram of anthracene

The phase diagram of anthracene was plotted using the Clausius-Clapeyron equation (Roux *et al.*, 2008),

$$\ln\frac{p_1}{p_2} = \frac{\Delta_x H}{R}\left(\frac{1}{T_2} - \frac{1}{T_1}\right) \tag{S2}$$

where $\Delta_x H$ was either the enthalpy of vapourization $\Delta_{vap}H$ or sublimation $\Delta_{sub}H$. These were taken as 79.9 kJ/mol and 101.9 kJ/mol, respectively, following Roux *et al.* (2008). Note that $\Delta_{sub}H$ is equal to the sum of $\Delta_{melting}H$ and $\Delta_{vap}H$. Finally, $R$ is the ideal gas constant.

The Clausius-Clapeyron equation requires knowledge of a reference vapour pressure, which we obtained using the Antoine equation $log_{10}p = A - (B/(T+C))$, where $p$ is the vapour pressure in bar at temperature $T$ in Kelvin. We used $\{A, B, C\} = \{4.72997, 2759.53, -30.753\}$ from Burgess (2020) to estimate anthracene's vapour pressure at the bulk melting point of 489 K (216 °C). We then extrapolated this vapour pressure across the desired temperature range using the Clausius-Clapeyron equation and plotted the ratios of the calculated $p$. Though these coefficients are valid only down to 496.4 K (Burgess, 2020), our approach ensured overlap of the vapour pressures at the melting point. We did not find any literature values of $p$ which resulted in a correct prediction of the melting point, which we attributed to the experimental difficulty in measuring the extremely low vapour pressures of solid anthracene (Roux *et al.*, 2008).

**Reference:** Donald R. Burgess, Jr. "Thermochemical Data" in NIST Chemistry WebBook, NIST Standard Reference Database Number 69, Eds. P.J. Linstrom and W.G. Mallard, National Institute of Standards and Technology, Gaithersburg MD, 20899, https://doi.org/10.18434/T4D303, (retrieved October 29, 2020).

#### S1.2.1 Phase diagram for designing restructuring experiments

Our anthracene experiments were designed based on the phase diagram shown in Figure 2a. The figure plots $S = p/p_0$, the saturation ratio with respect to liquid anthracene (Section 2), against temperature. By definition, $S = 1$ for bulk liquid anthracene, while $S < 1$ for capillary liquids due to their curvature. The figure illustrates capillary curvature for equivalent diameters of curvature $> 2$ nm (Ferry *et al.*, 2002).

For solid anthracene, which exists only below the bulk melting point $T_{\text{melt, bulk}}$, $S < 1$ due to the greater order of the solid phase. $T_{\text{melt, bulk}}$ is illustrated by the vertical line at 488 K.

The anthracene phase diagram also includes a second vertical line to illustrate the melting point of nanoscale amounts of anthracene. This melting point, $T_{\text{melt, p}}$, is substantially lower than $T_{\text{melt, bulk}}$ due to the well-understood effects of melting-point depression (Pawlow, 1909; Schmidt *et al.*, 1998; Nanda, 2009; Christenson, 2001) which have even been specifically demonstrated in the context of soot compaction (Chen *et al.*, 2016).

The magnitude of nanoscale melting-point depression is quantified by the material constant $\beta$, which has been calculated as 1.303 for anthracene using molecular dynamics simulations (Chen *et al.*, 2014). The results of this prediction are shown by the upper axis of Figure 2a, which shows the size of a spherical particle with a melting point equal to the lower axis, $T_{\text{melt, p}}$. The melting point function $T_{\text{melt}}(m_p)$ increases asymptotically towards the bulk melting point (215 °C) with particle size. Particles with $d_e$ greater than about 100 nm diameter have $T_{\text{melt}}(m_p)$ close to the bulk value, while small particles have much lower $T_{\text{melt}}(m_p)$ For example, $T_{\text{melt, p}} = 2$ nm is found at 348 K (75 °C), about 29% lower than the bulk melting point of $T_{\text{melt, bulk}} = 488$ K (215 °C). These values should be taken as qualitative, as the calculated $T_{\text{melt, p}}$ represents the equilibrium melting point; solids in confinement may remain liquid at even higher temperatures (Christenson, 2001). For example, Lopatkin *et al.* (1977) observed that anthracene films of thickness 10 nm to 16 $\mu$m form as liquids below 328 K. Nevertheless, these temperatures are comparable to our heater temperatures of 428–483 K, while some earlier work used lower temperatures (see Figures S2 and 6, and Section 4.3). Further details of $T_{\text{melt}}(m_p)$ and its calculation are given in Figure S2 and Section S1.4.

Figure 2b illustrates the use of this phase diagram to design the soot compaction experiments in Section 4. Each arc illustrates a process which we verified experimentally, primarily by measuring shape factors $\chi$ (Equation S1) as discussed below. The arcs illustrate capillary condensation at a single $S$ (single pore size) for simplicity; in reality capillary adsorption would resemble an isotherm (e.g. Ferry *et al.*, 2002), though with relatively less adsorption than normal, since our experiments involved short timescales and therefore allowed less time for capillary condensation, $t_c$ (Section 2). Note that, for simplicity, the arcs on Figure 2 do not emphasize the potential influence of freezing within capillary pores (Marcolli, 2014), which may lower $S$ above the particle after freezing.

Each of the phase changes in Figure 2 may be experimentally achieved as follows. Liquid condensation or solid deposition may be achieved by passing soot particles through a heated reservoir of a given coating material, followed by a cooling section. The cooling section will produce a profile of supersaturation that is dependent on the temperature profile across the heated and cooling sections, which depends on flow rate, tubing material, and tubing insulation (Nguyen *et al.*, 1987). The rate of increase of supersaturation will dictate the amount of time available for heterogeneous nucleation of vapours onto the particles (discussed in Section 2). The maximum attained supersaturation will dictate the amount of coating condensed onto the sample particles. Condensation may occur via one of two processes, depending on the soot–coating contact angle $\theta$ and the supersaturation reached: it may form either menisci (first step of CC in Figure 1) or beads (first step of NDAC in Figure 1) on the soot surface (Section 3). Competitive condensation to the walls of the system will also occur (Nguyen *et al.*, 1987).

Evaporation may be achieved by decreasing the saturation ratio of the aerosol. This may be achieved by removing vapours and is dramatically accelerated at elevated temperatures. For oleic acid coatings, where the melting point is above room temperature, evaporation is trivial and can be achieved with a catalytic stripper or thermodenuder. For anthracene coatings, the melting point is well above room temperature even after taking melting-point depression into account. Evaporation requires heating the anthracene coating above its melting point ($\geq 80$ °C) without allowing the coating to sublimate. We achieved this by passing anthracene-coated soot through our second coating apparatus, while the second apparatus was heated to high temperatures and also filled with anthracene. The sample then follows the evaporation curve in Figure 2 (thin blue solid line). Our apparatus was only able to sustain these conditions for a short period, before the anthracene in apparatus 2 was 'distilled' from the heating section to the cooler tubing downstream by sublimation-deposition.

When the second apparatus was depleted of anthracene, the sample followed the sublimation curve in Figure 2 (thin red line). Sublimation was also achieved by heating anthracene in the catalytic stripper and thermodenuder.

Thus, by combining the experimental conditions required to achieve each phase change, the four restructuring pathways in Table 1 may be achieved. To summarize, we achieved Path **LL** using oleic acid, and Paths **LS**, **SS**, and **SL** using anthracene. We achieved Path **LL** under conditions where mechanisms CC and BCC were possible, but not under conditions where NA was possible (Figure 1). The NA mechanism has been achieved previously by others ((Ma *et al.*, 2013a; Chen *et al.*, 2018)) as discussed in Section 3.

## S1.3 Additional Experimental Methods

### S1.3.1 Soot samples

Soot was generated by a miniCAST 5201c soot generator (Jing Ltd., Switzerland) which consists of a partially-quenched propane diffusion flame. The miniCAST was operated under lean combustion combustions, with the following settings for each {gas, flow in mL min$^{-1}$}: {propane, 60}, {premixed nitrogen, 0}, {combustion air, 1550}, {dilution air, 2e4}. Premixed nitrogen refers to potential premixing with combustion air. We avoid using the CAST burner to produce so-called "low-BC fraction" soot, which is quenched earlier in the combustion process to produce incompletely-graphitized particles that are poor surrogates for atmospheric soot (Maricq, 2014). A 1.5 m$^3$ steel tank was used as a reservoir during the experiments.

### S1.3.2 Real-time measurements

As depicted in Figure S1, a number of instruments were used to characterize the soot samples online. The total flow was 0.71 min$^{-1}$. A first Differential Mobility Analyzer (DMA1, built at PSI, with similar dimensions to the Model 3801 DMA of TSI Inc., USA) pre-selected particles, typically at 300 nm. The resulting monodisperse aerosol was either sampled directly or processed as described below.

The aerosol was measured by a Scanning Mobility Particle Sizer (SMPS; comprised of a similar DMA, DMA2, plus a TSI 3022a Condensation Particle Counter, CPC), scanning from 100 to 700 nm. In some cases, we bypassed DMA2 in order to test whether new particles (with diameter $<100$ nm) were forming in the coating apparatus; such conditions were always avoided. In parallel, a Single Particle Soot Photometer (SP2; Schwarz *et al.*, 2006; Laborde *et al.*, 2012) simultaneously measured single-particle BC mass $m_{BC}$ by laser-induced incandescence and optical size by light scattering. The SP2 was frequently operated behind an Aerosol Particle Mass analyzer (APM; Model 3601, Kanomax Japan), which classifies particles by their mass-to-charge ratio $m/q$. Using custom software written in LabView, the APM was scanned through $m/q$ settings while the SP2 was used to selectively count singly charged ($q = 1$) BC particles exiting the APM. Like a high-resolution optical particle counter, the SP2 is able to discriminate singly charged particles because its signals are proportional to particle mass, which reduces ambiguity in the data relative to APM-CPC measurements. The modes of the resulting mass distributions were fitted by lognormal or asymmetric normal curves (Tajima *et al.*, 2011) and used to define the total particle mass.

The particle mass measurements were converted to particle volume by

$$V_1 = (V_0 + V_{OM}) = \frac{m_0}{\rho_{BC}} + \frac{(m_1 - m_0)}{\rho_{OM}} \tag{S3}$$

where $m_0$ and $m_1 = (m_0 + m_{OM})$ are the mass of unprocessed or processed soot particles, $\rho_{BC} = 1.8$ kg m$^{-3}$ is the material density of BC and $\rho_{OM}$ is the material density of condensed or deposited organic matter. For unprocessed soot ($m_{OM} = 0$), $V_1 = V_0$.

Shape factors were calculated by iteratively solving Equation S1 above, with $d_{mob}$ defined by the setpoint of DMA1. In calculating $C_c$, we used the mean free path in air of 66.35 nm given by Jennings (1988).

### S1.3.3 Offline measurements

TEM was performed with a JEOL 2010 at 200 keV after electrophoretic sampling particles onto 300-mesh copper grids with a 1-2 nm amorphous carbon film. The copper grids were obtained from Quantifoil (Großlöbichau, Germany).

Sampling was performed using a Partector (Naneos GmbH, Windisch, Switzerland), which charges particles with a corona charger before electrophoretic sampling. TEM samples were performed for the maximum coating volumes (maximum $V_1/V_0$).

### S1.3.4 Soot coating addition and removal

Soot coatings were added using one of two apparatuses. The first was similar to that described by Moteki and Kondo (2007) and was constructed at PSI by M. Heringa. It consisted of two concentric stainless steel tubes (12 mm and 8 mm). Aerosol samples flowed in through the outer tube and out through the inner tube. The outer tube was terminated at one end by a stainless steel cap, which was filled with either oleic acid or anthracene crystals. The cap was placed on a hot plate and heated to saturate the aerosol flow. The second apparatus consisted of a simple copper tube wrapped first in heating tape, then in ice–water packs and was similar to that described by (Nguyen *et al.*, 1987).

Coating materials were placed in a bend in the copper tube. The second apparatus was able to reach higher coating thicknesses due to the active cooling stage and the high thermal conductivity of copper. The second apparatus was also necessary for the anthracene evaporation experiments described in Section 4.1.

The coating apparatuses were operated at different temperatures for oleic acid (90 %, Sigma-Aldrich) and anthracene (99 %, Sigma-Aldrich). For oleic acid condensation, the heater was operated between 94 and 140 °C. Above 140 °C new-particle formation occurred, indicating that excessive supersaturations were being reached after the coating apparatus. Anthracene coatings were obtained with a heater temperature of $95-202$ °C. Above 202 °C, the reservoir emptied too quickly for the measurements. Note that the thermocouple was placed between the heating plate and the sample, so the reported temperatures are upper limits. The lower half of the apparatus was insulated and the upper half cooled by room air to maximize the achieved supersaturations. For some anthracene experiments, even higher supersaturations were achieved by immersing the lower third of the apparatus in oil. The oleic acid experiments lasted one day and were performed in sequence: the highest temperatures were used last. Afterwards, the oleic acid was observed to have transformed from a light-yellow to a reddish-brown liquid, indicating that a chemical transformation had taken place. However, our results are consistent with previous work (Ghazi and Olfert, 2013) and indicate that this change had a negligible effect on the density of the oleic acid which was taken as 895 kg m$^3$ in our calculations.

The anthracene experiments lasted three days in total and were not performed in sequence. The maximum temperature used was 202 °C. In the early minutes of some high-temperature anthracene experiments, the SMPS appeared to measure the tail of a nucleation mode. As no difference in the results was observed when this mode was present, we conclude that it did not influence the experiments.

The primary denuder used in this work was a catalytic stripper (CS015, Catalytic Instruments GmbH, Rosenheim, Germany), which vapourizes and then oxidizes organic molecules at 350 °C. In some experiments, this stripper was replaced with the activated-charcoal thermodenuder described by Burtscher et al. (2001).

## S1.4  Melting-point depression: further discussion and calculations

Physically, melting point depression may be alternatively viewed a reduction of intermolecular forces due to the disorder of nanoscopic quantities of anthracene (Aubin and Abbatt, 2006) or as the result of balancing bulk and surface energy terms (Christenson, 2001). The magnitude of melting-point depression for anthracene was estimated for Figure S2 using a size-dependent thermodynamic model of nanoscopic melting-point depression (Nanda, 2009):

$$\frac{T_{m,\,\text{particle}}}{T_{m,\,\text{bulk}}} = 1 - \left(\frac{\beta}{d_e}\right) \tag{S4}$$

where $T_{m,\,\text{particle}}$ is the melting point of a particle with spherical-equivalent diameter $d_e$ and $T_{m,\,\text{bulk}}$ is the melting point of bulk material ($d_e >$ 100 nm, see Figure S2). This model has been validated experimentally (Nanda, 2009) and is valid for materials which melt homogeneously or from the surface inwards. Using a molecular-dynamics model, Chen et al. (2014) found that pyrene melts via both of these mechanisms (Chen et al., 2014) and estimated the material-specific parameter $\beta$ as 1.303 for pyrene (four compact aromatic rings) and 1.312 for coronene (seven compact rings). As these two values are virtually identical we have used $1.3 \pm 25\%$ for anthracene (three linear rings)in Figure S2. This $\beta$ results in a melting point for 1 nm spheres which is about 2/3 of the melting point of bulk anthracene, which is close to the melting point depression measured experimentally by Lopatkin et al. (1977) for anthracene thin films. This consistency is in line with expectations, since $\beta$ is a thermodynamic parameter, a function of the solid density, liquid density, surface energies of the solid–vapour and liquid–vapour interfaces, and the bulk latent heat of fusion (Nanda, 2009).

We note that our use of Equation S4 is different to that of Chen et al. (2016), who were the first to consider the effects of melting-point depression on soot restructuring. Those authors considered thin-film thicknesses when applying Equation S4, whereas we have considered equivalent spheres ($d_{\text{sph-equiv}}$). They observed PAH restructuring at equivalent film thicknesses (calculated from single-particle mass measurements) of 0.2 nm, which is half the 0.4 nm thickness of a graphene sheet (Nemes-Incze et al., 2008). They therefore argued that their observations demonstrated capillary condensation and restructuring, in agreement with the conclusions of the present work..

## S2  Supplementary figures

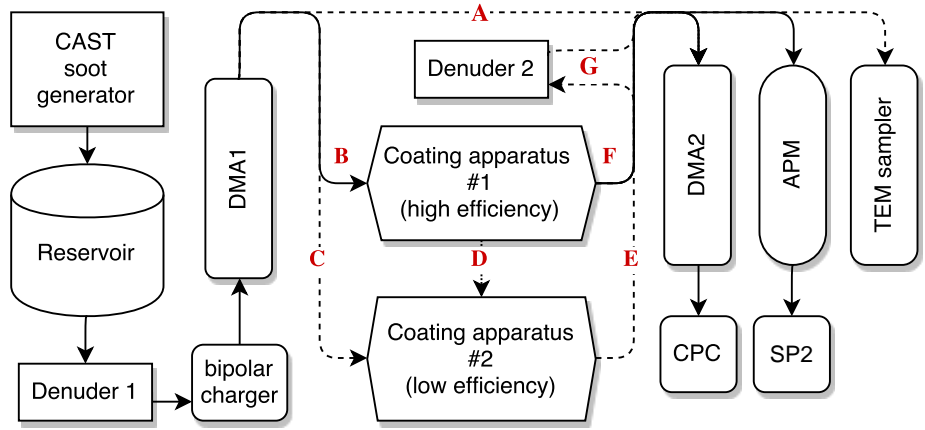

**Figure S1.** Apparatus used to process soot particles through the different pathways given in Table 1 and related Figure 3. Pathways were achieved as follows. Path (**LL**): route **CEG** with oleic acid; Path (**LS**): route **CEG** with anthracene; Path (**SS**): route **BFG** with anthracene; Path (**SL**): route **BDE** with anthracene and anthracene-saturated walls in the second coating apparatus. Reference untreated samples were measured through line **A**. The denuder **G** was removed to measure coated particles. Dashed lines indicate temporary connections. The SP2 was used to differentiate particle charges only, not to measure single-particle mass.

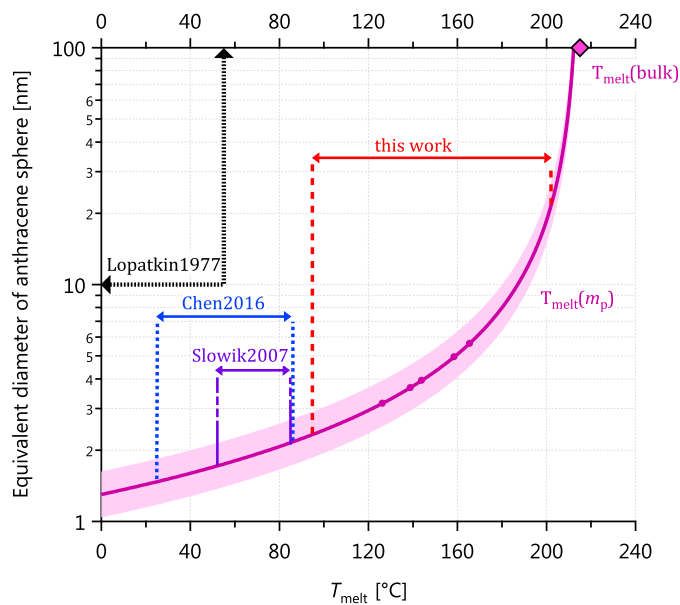

**Figure S2.** Melting point depression of nanoscopic quantities of anthracene, as discussed in the text. Overlaid are the temperatures at which these and related anthracene experiments have been carried out. The curve indicates the melting point depression for nanoscopic quantities of anthracene based on molecular dynamics simulations at the points marked with spheres (Chen *et al.*, 2014). The asymptote of this curve is the melting point of bulk anthracene (216 °C, ◆). The studies of Chen *et al.* (2016) and Slowik *et al.* (2007) are similar to ours and discussed in Section 4.3; the study by (Lopatkin *et al.*, 1977) showed that anthracene thin films (thickness 10 nm to 16 μm) form as liquids below 55 °C.

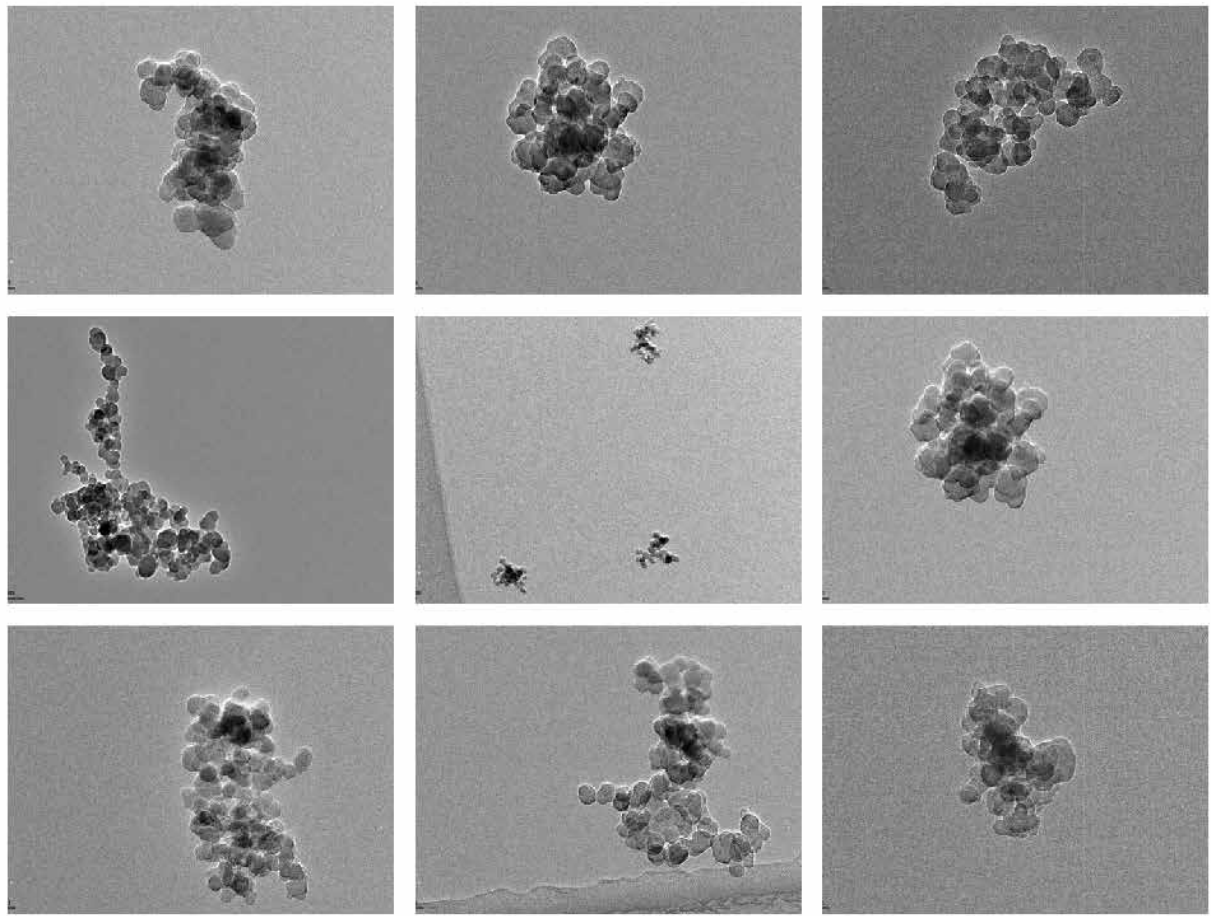

**Figure S3.** TEM images of 300 nm soot after anthracene coating ($m_1/m_0 = 2.6$) and denuding via Path **1** of Figure 3. Scale bars are omitted since all soot was composed of monomers with 30 nm diameter (Figure S6).

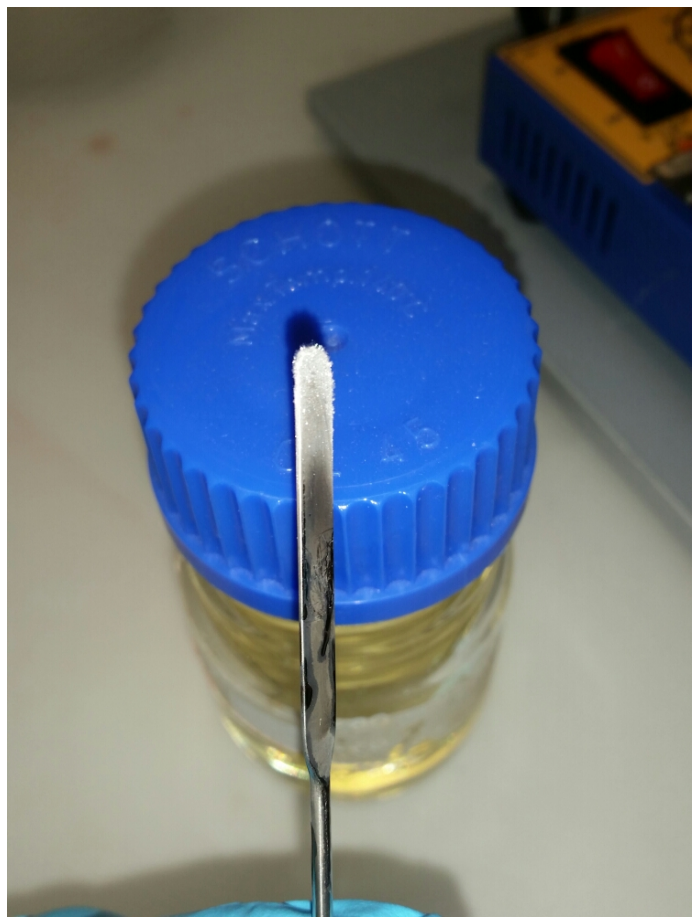

**Figure S4.** Photograph of a spatula upon which anthracene was deposited in our coating apparatus. The crystalline structure of the deposit is clearly visible. The conditions used for this experiment were used for the experiment labelled Path **3** in Figure 3.

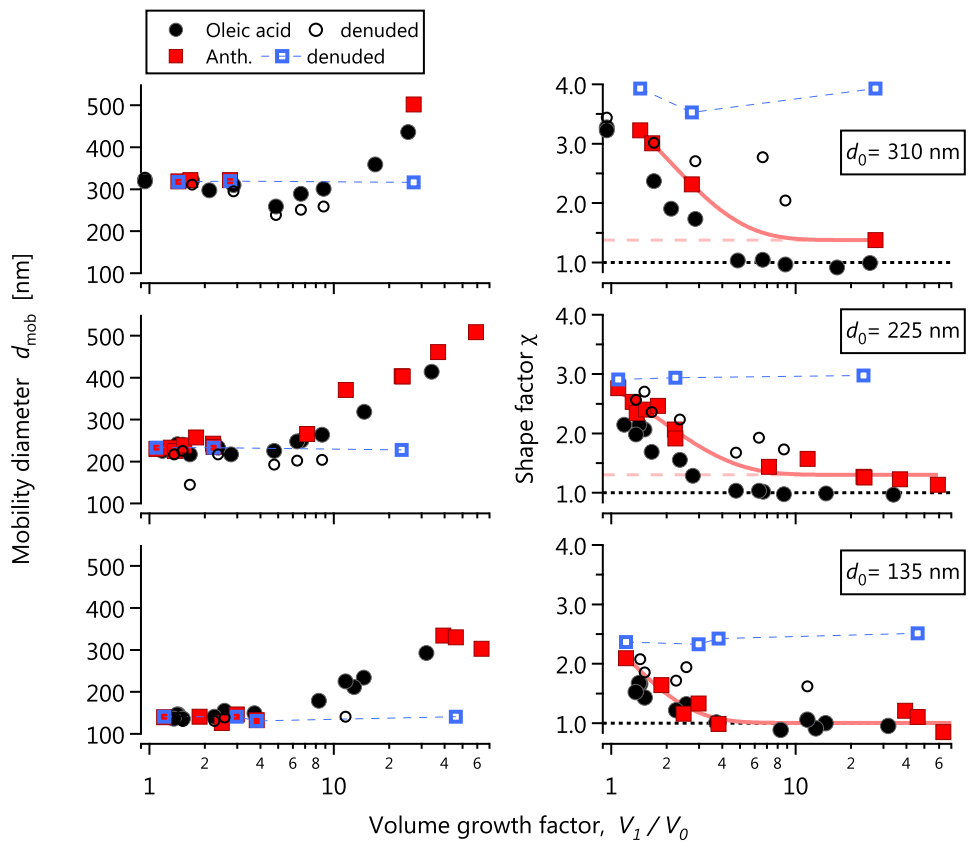

**Figure S5.** Data from Slowik *et al.* (2007), reanalyzed to illustrate coating/denuding without a change in the final $d_{mob}$ or $\chi$. Initial particle diameters are 135, 225, and 310 nm. This demonstrates deposition/sublimation of anthracene (Path **3** in Figure 3). Left and right panels show mobility diameter and shape factor $\chi$ as a function of coating volume. Circles reflect oleic acid condensation, squares anthracene deposition. Since $\chi \approx 1$ at large $V_1/V_0$ for both substances, anthracene likely condensed as a supercooled liquid. The fits on the right hand side reach a minimum of $\chi = 1.37$, 1.3, and 1.0 for the 310, 225, and 135 nm particles, respectively.

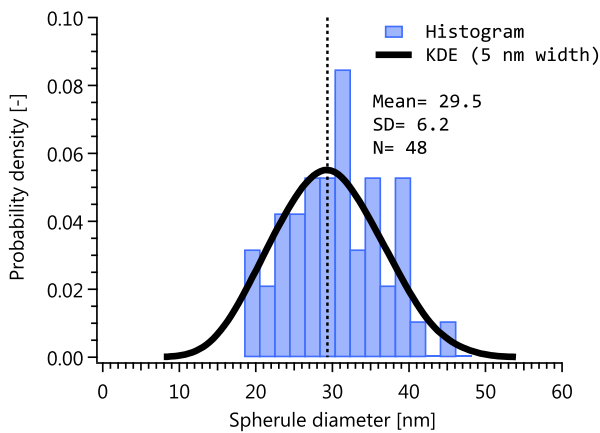

**Figure S6.** CAST Soot spherule diameters, measured manually from the TEM images. KDE is the kernel density estimated with a Gaussian kernel.